\documentstyle[aps,prb,twocolumn]{revtex}

\title{Critical Currents of Josephson-Coupled Wire Arrays}
\author{J.~Kent~Harbaugh and D.~Stroud}
\date{\today}

\newcommand{\ba}{\begin{eqnarray}}
\newcommand{\ea}{\end{eqnarray}}

\begin{document}

\maketitle

	\abstract{We calculate the current-voltage characteristics and
critical current $I_c^{\mbox{\tiny array}}$ of an array of
Josephson-coupled superconducting wires.  The array has two layers, each
consisting of a set of parallel wires, arranged at right angles, such that
an overdamped resistively-shunted junction forms wherever two wires cross. 
A uniform magnetic field equal to $f$ flux quanta per plaquette is applied
perpendicular to the layers.  If $f = p/q$, where $p$ and $q$ are mutually
prime integers, $I_c^{\mbox {\tiny array}}(f)$ is found to have sharp
peaks when $q$ is a small integer.  To an excellent approximation, it is
found in a square array of $n^2$ plaquettes, that $I_c^{\mbox{\tiny
array}}(f) \propto (n/q)^{1/2}$ for sufficiently large $n$.  This result
is interpreted in terms of the commensurability between the array and the
assumed $q \times q$ unit cell of the ground state vortex lattice. 

\section{Introduction} \label{sec:intro}

	There has been much recent interest in Josephson junction arrays
with so-called long-range interactions.  Such arrays typically consist of
two layers of $N$ parallel superconducting wires, arranged at right
angles, in such a fashion that a Josephson junction is formed at each
point where two wires cross (see Fig.~\ref{fig:geom}).  In this geometry,
each wire in one layer is Josephson-coupled to all the wires in the other
layer.  If $N \gg 1$, each wire has many ``nearest-neighbor'' wires to
which it is coupled.  Under these conditions, the thermodynamic properties
of the array are accurately described by mean-field theory.  Indeed, it
has been shown\cite{vinokur} that the usual mean-field theory for phase
transitions in such arrays in a transverse magnetic field\cite{shih}
becomes exact in the limit $N \rightarrow \infty$, provided that each wire
can be considered as a locus of constant superconducting phase. 
Experiments on finite arrays ($N \approx 17$) have recently confirmed the
accuracy of mean-field predictions.\cite{tinkham1} For very large arrays,
the assumption that each wire is a locus of constant phase breaks down,
and the mean-field theory has to be corrected to take account of the phase
variation along the array.\cite{harbaugh1} This same effect causes the
critical current of the array to saturate at a finite value which depends
on the Josephson penetration depth, even as $N \rightarrow
\infty$.\cite{tinkham2,harbaugh2}

	The exactness of mean-field theory is believed to have an analog
in the {\em dynamics} of these arrays.  Chandra {\it et
al.}\cite{chandra1,chandra2} have studied a version of local dynamics
equivalent, in the large-N limit, to the resistively-shunted junction
(RSJ) model with Gaussian thermal noise and no applied current (they use a
different numerical approach from that followed here).  In the dynamical
case, the effects of magnetic-field-induced frustration combine with those
arising from the fact that each wire has many other ``nearest-neighbor''
wires with which it interacts.  The result is a kind of glassy behavior,
with many metastable states separated by large energy barriers, critical
slowing down near the phase transition, and typically glassy aging
effects. 

	In this paper, we extend the RSJ model described above to include
applied currents.  Specifically, we calculate the current-voltage ($IV$)
characteristics and the critical currents $I_c^{\mbox {\tiny array}}$ for
arrays of crossed Josephson junctions, in typical experimental geometries
as shown in Fig.~\ref{fig:geom}.  We assume that, even though $N \gg 1$,
each wire can be taken as a locus of constant phase, and we study how
$I^{\mbox{\tiny array}}_c$ is affected by a transverse magnetic field of
magnitude $f$ flux quanta per plaquette.  Our principal result is that
$I^{\mbox{\tiny array}}_c(f)$ is extremely sensitive to $f$, in a manner
related to the commensurability between the lattice and the vortex
structure of the induced array supercurrents.  A similar sensitivity is
found in the array critical temperature $T_c(f)$, as calculated by
mean-field theory,\cite{tinkham2} and as confirmed in detail by
experiments.  It would be of great interest if similar experiments could
be carried out to confirm our calculated $I^{\mbox{\tiny array}}_c(f)$. 

	We turn now to the body of the paper.  In Section~\ref{sec:model}
we describe the geometry of our calculations and the model used to
calculate the $IV$ characteristics and $I_c^{\mbox {\tiny array}}$.  Our
numerical results are given in Section~\ref{sec:results}, followed by a
brief discussion in Section~\ref{sec:discuss}.

\section{Model} \label{sec:model}

	As shown in Fig.~\ref{fig:geom}, the array consists of two layers
of parallel superconducting wires arranged at right angles.  A resistively
shunted Josephson junction is assumed to form at each point where two
wires cross.  To determine $I_c^{\mbox {\tiny array}}$, we assume that a
constant bias current $I_B$ is injected into one wire and extracted from a
different wire.  The $IV$ characteristics are determined by calculating
the time-averaged voltage drop across these two wires. We consider two
configurations for current injection.  In the first, the current is
injected into a wire at the left edge of one layer and extracted from the
wire at the right edge of the same layer (left-hand side of
Fig.~\ref{fig:geom}).  In the other configuration, the input and output
wires are centered in different layers of the array (cf.\ right-hand side
of Fig.~\ref{fig:geom}). We also assume that a time-independent, uniform
magnetic field $B$ is applied perpendicular to the array, of magnitude $B
= f\Phi_0/a^2$, where $\Phi_0 = hc/(2e)$ is the flux quantum, and $a^2$ is
the area of each square plaquette. 

     We assume that the phase of the superconducting order parameter is a
constant on each wire, independent of position along the
wire.\cite{comment1} With this assumption, we can write down the equations
determining the flow of current through the system.  We assume an array of
$N \times M$ wires, and model the intersections between the crossed wires
as overdamped, resistively-shunted Josephson junctions.  The equations
determining the time-variation of phases on each wire can then be written
down using Kirchhoff's Laws.  For example, in the perpendicular current
input configuration, the equations take the form
  \ba
  I_B\delta_{jm} &=& \sum_{k=1}^N \left[ I_c \sin(\psi_j-\phi_k-A_{kj}) +
I_{kj} + I_{L;kj} \right], \nonumber\\
  I_B\delta_{kn} &=& \sum_{j=1}^M \left[ I_c \sin(\psi_j-\phi_k-A_{kj}) +
I_{kj} + I_{L;kj} \right]. 
  \label{eq:kirch}
  \ea
  Here $I_B$ is the constant bias current, which is injected into wire $m$
in one layer and extracted from wire $n$ in the other layer; $\phi_k$ is
the phase of the $k$th wire in one layer; and $\psi_j$ is that of the
$j$th wire in the other layer.  $I_c$ denotes the critical current of a
junction formed at the intersection of any two wires (the junctions are
assumed to be identical).  The supercurrent flowing through the $ij$th
junction is then $I_c\sin(\psi_j-\phi_k-A_{kj})$, where the quantity in
parentheses is the gauge-invariant phase difference across the junction. 
We use a gauge such that the phase factor $A_{kj}$ arising from the vector
potential of the magnetic field is $A_{kj} =2\pi kj
Ba^2/\Phi_0$.\cite{comment2} $I_{kj}$ denotes the normal current through
the shunt resistance $R$ in the junction at coordinates $(ka, ja)$; it is
given by
  \ba I_{kj} = \frac{\hbar}{2eR} \frac{d}{dt} (\psi_j - \phi_k - A_{kj}). 
  \ea
  Finally, at finite temperatures, one must also include a Langevin noise
current $I_{L;kj}$ in parallel to the other currents. In the present
calculations, we assume temperature $T = 0$, so that $I_{L;kj}$
vanishes.\cite{chung} The equations describing the parallel current
injection can be written down in an analogous fashion. 

	Eqs.\ (\ref{eq:kirch}) can be conveniently expressed in a
dimensionless form, such that current is measured in units of $I_c$, and
time in units of $\hbar/(2eI_cR)$, and then rewritten in a fashion which
is more convenient for numerical solution.  First, we rearrange
Eqs.~(\ref{eq:kirch}) to obtain
  \ba
  \dot{\phi}_k &=& \frac{1}{N} \sum_{i=1}^N \dot{\phi}_i + \frac{1}{NM}
\left[ i_B - \sum_{i=1}^N \sum_{j=1}^M \sin(\psi_j - \phi_i - A_{ij})
\right] \nonumber\\ && - \frac{1}{M} \left[ i_B\delta_{kn} - \sum_{j=1}^M
\sin(\psi_j - \phi_k - A_{kj}) \right], \nonumber\\
  \dot{\psi}_k &=& \frac{1}{N} \sum_{i=1}^N \dot{\phi}_i + \frac{1}{N}
\left[ i_B\delta_{km} - \sum_{i=1}^N \sin(\psi_k - \phi_i - A_{ik}
\right]. 
  \label{eq:rerng}
  \ea
  Of the $N+M$ degrees of freedom in these equations, one may be chosen
freely.  We fix this degree of freedom by arbitrarily choosing
  \ba
  \frac{1}{N} \sum_{i=1}^N \dot{\phi}_i &=& -\frac{1}{2NM} \left[ i_B -
\sum_{i=1}^N \sum_{j=1}^M [\sin(\psi_j - \phi_i - A_{ij}) \right]. 
  \ea
  With this choice, and making use of the double-angle formulas for the
sines and cosines, Eqs.~(\ref{eq:rerng}) take the final form
  \ba
  -\dot{\phi}_k &=& -\frac{1}{2NM} \left[ i_B - \sum_{i=1}^N \sum_{j=1}^M
\sin(\psi_j - \phi_i - A_{ij}) \right] \nonumber\\ && + \frac{1}{M} \left[
i_B\delta_{kn} - \sum_{j=1}^M \sin(\psi_j - \phi_k - A_{kj}) \right],
\nonumber\\
  \dot{\psi}_k &=& -\frac{1}{2NM} \left[ i_B - \sum_{i=1}^N \sum_{j=1}^M
\sin(\psi_j - \phi_i - A_{ij}) \right] \nonumber\\ && + \frac{1}{N} \left[
i_B\delta_{km} - \sum_{i=1}^N \sin(\psi_k - \phi_i - A_{ik}) \right]. 
  \label{eq:final}
  \ea
  The form of these equations is quite suggestive.  Namely, the
time-variation of phases in {\em one} layer depends only on certain
averages over phases in the {\em other} layer.  A similar result is known
to hold for the averages of certain {\em equilibrium} functions of the
phases in each layer, in the thermodynamic (i.e.\ large N and large M)
limits. A consequence of this result in the equilibrium case is that
mean-field theory is exactly valid for the thermodynamics of this system
in the limit of large N and large M.\cite{vinokur}

	Although we have not found an analytical solution to these
dynamical equations, they are easily solved numerically for any $N$ and
$M$ using a standard Runge-Kutta method.  We have used a fixed-step-size,
fourth-order Runge-Kutta algorithm, with time steps ranging between 0.01
and 0.1 dimensionless time units.  We obtained the d.c.\ voltage $V$ by
averaging the instantaneous voltage over 100 time units.  In order to
obtain the $IV$ characteristics, the input current was slowly swept from
high to low currents.\cite{hysteresis} After each current step, the system
was given 100 time units to reach steady state before the time-averaged
voltage was computed.  Zero-temperature calculations were carried out for
two different $N \times N$ array sizes, $N =18$ and $N = 14$, for each
input configuration. 

\section{Results} \label{sec:results}

	Figure~\ref{fig:iv} shows the calculated $IV$ curves for several
fields in the {\em parallel} current configuration.  The field is
expressed in terms of the {\em frustration} $f = Ba^2/\Phi_0$, the number
of flux quanta per plaquette.  By definition, the array critical current
$I^{\mbox{\tiny array}}_c$ is the largest bias current for which the
time-averaged voltage $\langle V \rangle_t= 0$.  By inspection of
Fig.~\ref{fig:iv}, $I^{\mbox{\tiny array}}_c$ is considerably suppressed
by application of a finite magnetic field.  If $f$ is a rational fraction
denoted $f = p/q$ where $p$ and $q$ are integers with no common factors,
then the figure also shows that $I_c^{\mbox{\tiny array}}(f)$ is
suppressed more at frustrations with larger values of $q$. 
 
	These observations are made more quantitative in
Figs.~\ref{fig:icfd} and~\ref{fig:icfx}, which show the calculated
$I^{\mbox{\tiny array}}_c(f)$ for a wide range of fields. 
Figure~\ref{fig:icfd} shows $I^{\mbox{\tiny array}}_c(f)$ for an $18
\times 18$ array in both the perpendicular (top) and parallel (bottom)
input current configurations.  In this case, $I^{\mbox{\tiny array}}_c(f)$
was obtained by directly calculating the voltage across the array for a
range of bias currents and recording the highest current with a zero mean
voltage.  In order to reduce the numerical noise, we have smoothed the
critical current data as a function of frustration with a Savitzky-Golay
filter.\cite{filter} Clearly, $I^{\mbox{\tiny array}}_c(f)$ has sharp
peaks near fractions $f = p/q$ with small denominators $q$. 

	Since it can be difficult to determine numerically the precise
current at which $\langle V \rangle _t \rightarrow 0$, we have also
calculated $I^{\mbox{\tiny array}}_c(f)$ by extrapolating the $IV$ curve
above $I_c^{\mbox{\tiny array}}$, assuming the functional form
  \ba
  {\left[ \frac{V}{I_cR} \right]}^2 &=& m^2 \left[ {\left( \frac{I_B}{I_c}
\right)}^2 - {\left( \frac{I^{\mbox{\tiny array}}_c}{I_c} \right)}^2
\right],
  \label{eq:fit}
  \ea
  and determining $m$ and $I^{\mbox{\tiny array}}_c$ by the least-squares
fitting procedure.  The resulting data were also smoothed using the same
Savitzky-Golay filter.  This extrapolation algorithm, used to calculate
the data in Fig.~\ref{fig:icfx}, considerably reduces the numerical noise
near $I^{\mbox{\tiny array}}_c$, and the smaller peaks in $I^{\mbox{\tiny
array}}_c(f)$ stand out more clearly than in the direct calculation. 
Using this procedure, we have calculated $I^{\mbox{\tiny array}}_c(f)$ for
both an $18 \times 18$ (top) and a $14 \times 14$ (bottom) array in the
{\em parallel} current injection configuration.  Clearly, there are peaks
in $I^{\mbox{\tiny array}}_c(f)$ at all fractions $f = p/q$ such that $q
\leq 16$ in the larger array or $q \leq 12$ in the smaller array.  We now
discuss some possible reasons for this behavior. 

\section{Discussion} \label{sec:discuss}

	The present results show that $I^{\mbox{\tiny array}}_c(f)$ is a
sensitive function of the applied field $f$. At $f = 0$, the array is
unfrustrated and $I^{\mbox{\tiny array}}_c(f)$ is maximized.  At a
rational frustration $f = p/q$ where $p$ and $q$ are integers with no
common factor and $q \leq N-2$, our numerical results suggest that height
of the peaks in $I^{\mbox{\tiny array}}_c(f)$ approximately obeys the
relation
  \ba
  i^{\mbox{\tiny array}}_c \sim r^{1/4} (s/P^2)^{1/2}. 
  \label{eq:peaks1}
  \ea
  Here $r$ is the number of flux lattice unit cells (assumed to be of size
$q \times q$) which can be inscribed in the array, $s = rq^2$ is the
number of array plaquettes which are contained in all $r$ of the $q \times
q$ cells, and $P^2 = (N-1)^2$ is the total number of plaquettes in the
array ($N=M$ in a square array).  This form can be made explicit by
writing $r = ({\rm int}[P/q])^2$, where ${\rm int}[x]$ is the greatest
integer $n \leq x$. Then Eq.~\ref{eq:peaks1} can also be written as
  \ba
  i^{\mbox{\tiny array}}_c \sim (q/P) {\left( {\rm int}[P/q]
\right)}^{3/2}. 
  \label{eq:peaks2}
  \ea
  In the limit of large $N$, this result takes the even simpler form
  \ba
  i^{\mbox{\tiny array}}_c \sim (n/q)^{1/2},
  \label{eq:peaks3}
  \ea
  where $n = N-1$. The values predicted by eq.~\ref{eq:peaks1} are plotted
as crosses in Fig.~\ref{fig:icfx}.  Agreement with the numerical data is
remarkably good. 

	We can understand Eq.~\ref{eq:peaks1} as a measure of the
commensurability between the field and the array.  We expect that, just as
in more conventional Josephson arrays with square plaquettes, the ground
state energy configuration in a field $f = p/q$ and no applied current has
a flux lattice unit cell consisting of $q \times q$ plaquettes.  A {\em
finite} wire array at a given field can contain a certain number of such
cells, with a number of plaquettes left over.  We may expect that the
current required to depin such unit cells, and hence $I^{\mbox{\tiny
array}}_c(f)$, will increase as the number of flux lattice unit cells
which can fit into the array becomes larger, and the number of
``remainder'' plaquettes becomes smaller. Indeed our analytical
expression, which fits our numerical data, expresses $I^{\mbox{\tiny
array}}_c(f)$ in this fashion. Note also that, according to
Eq.~\ref{eq:peaks3}, for large values of $N$, $i^{\mbox{\tiny array}}_c
\sim q^{-1/2}$.  Thus, in this regime, $i^{\mbox{\tiny array}}_c(f)$
depends on $q$ in the same way as the mean-field transition $T_c(f)$
reported in Ref.~\cite{tinkham1}. The proportionality between
$i^{\mbox{\tiny array}}(f)$ and $T_c(f)$ is is the same as that which is
believed to hold between critical current and transition temperature in
more conventional two-dimensional Josephson junction
arrays.\cite{teitel83}

	Finally, we comment briefly on our results in light of the work of
Refs.~\cite{chandra1,chandra2}.  Their picture of frustrated arrays may
well hold true even in the presence of an applied current.  However, it is
difficult to draw any quantitative conclusions from our own calculations
about the energy barriers they discuss, other than those which are
suggested by the field-dependence of the critical current.  These results
obviously suggest that the barriers are smallest for largest values of
$q$, though the reasons for the analytical dependence ($\propto q^{-1/2}$)
remain a matter for speculation.

\section{Acknowledgments} \label{sec:acknw}

	We gratefully acknowledge support by the National Science
Foundation, through grant DMR97-31511.

\newpage

\begin{figure}
	\caption{Sketch of array geometry.  The array consists of two
layers of parallel wires, arranged at right angles. A resistively-shunted
Josephson junction forms at each wire intersection.  There are two
configurations for the current injection: {\em parallel} injection and
extraction wires (shown on left); and {\em perpendicular} injection and
extraction wires (on right).}
	\label{fig:geom}
\end{figure}

\begin{figure}
	\caption{Current-voltage curves at several frustrations for an $18
\times 18$ array with {\em parallel} input wire configuration.  $V$
denotes time-averaged voltage between input and output wires.}
	\label{fig:iv}
\end{figure}

\begin{figure}
	\caption{The critical current $I_c^{\mbox {\tiny array}}(f)$
plotted versus frustration $f$ for an $18 \times 18$ array in the {\em
perpendicular} input wire configuration (upper plot) and the {\em
parallel} configuration (lower plot).  These data are obtained from a
direct calculation, smoothed using the Savitzky-Golay
filter.\protect\cite{filter} The hash marks at top denote each rational
frustration $f = p/q$ with $q \leq 16$.  Frustrations with $p = 1$ are
denoted with bold hash marks.}
	\label{fig:icfd}
\end{figure}

\begin{figure}
	\caption{$I_c^{\mbox {\tiny array}}(f)$ for an $18 \times 18$ array
(upper plot) and a $14 \times 14$ array (lower plot) in the {\em parallel}
input wire configuration.  These data are calculated by a least-squares
fit to the functional form of Eq.~\ref{eq:fit}, then smoothed using the
Savitzky-Golay filter.\protect\cite{filter} Hash marks at top denote each
rational frustration $f = p/q$ with $q \leq 16$ (upper plot) or $q \leq
12$ (lower plot).  Frustrations with $p = 1$ have bold hash marks.  The
$\times$ symbols denote the peak heights as predicted by
Eq.~\ref{eq:peaks2}, using the same, arbitrary normalization for each
plot.}
	\label{fig:icfx}
\end{figure}


\begin{thebibliography}{}
\bibitem{comment1} In order for this condition to prevail, the array size
must be less than the Josephson screening length,\cite{harbaugh2} i.e.,
for a square array, $N \ll \sqrt{I^{\mbox{\tiny wire}}_c /I_c}$.  Of
course, even if this condition holds, the phase is constant along the wire
only if a particular gauge is chosen, as discussed below.
\bibitem{comment2} We choose a gauge such that the phase variables are
constant on each wire and the dependence on applied magnetic field is
contained entirely in the factors $A_{kj}$, as shown in
Eq.~(\ref{eq:kirch}).  See Ref.~\cite{harbaugh2} for a discussion of this
gauge in a slightly different context.
\bibitem{hysteresis} The calculated $IV$ curves appeared to be
non-hysteretic, except for small, apparently random fluctuations near
$I^{\mbox{\tiny array}}_c$.
\end{thebibliography}
\end{document}